\documentclass[prd,aps,showpacs,nofootinbib,tightenlines]{revtex4}
\usepackage{mathrsfs}
\usepackage{amsmath}
\usepackage{amssymb}
\usepackage{epsfig}
\usepackage{graphicx}
\usepackage{booktabs}
\usepackage{multirow}
\usepackage{subfigure}
\usepackage[section]{placeins}
\usepackage{float}
\begin{document}
\newcommand{\psl}{ p \hspace{-1.8truemm}/ }
\newcommand{\nsl}{ n \hspace{-2.2truemm}/ }
\newcommand{\vsl}{ v \hspace{-2.2truemm}/ }
\newcommand{\epsl}{\epsilon \hspace{-1.8truemm}/\,  }

\title{Nonleptonic two-body decays of $\Lambda_b\rightarrow \Lambda_c \pi, \Lambda_cK$ in the
perturbative QCD approach  }
\author{Chao-Qi Zhang$^1$}
\author{Jia-Ming Li$^1$}
\author{Meng-Kun Jia$^1$}
\author{Zhou Rui$^1$}\email[Corresponding  author: ]{jindui1127@126.com}
\affiliation{$^1$College of Sciences, North China University of Science and Technology,
                          Tangshan 063009,  China}
\date{\today}
\begin{abstract}
We study the color-allowed $\Lambda_b\rightarrow \Lambda_c \pi, \Lambda_cK$ decays in the perturbative QCD approach (PQCD)
to lowest order in strong coupling constant $\alpha_s$.
Both the factorizable and nonfactorizable contributions are taken into account in our calculations.
It is found that these processes are dominated by the factorizable contributions,
while the important nonfactorizable contributions can enhance the branching ratios by about 30$\%$.
The decay branching ratios are predicted to be $\mathcal{B}(\Lambda_b\rightarrow \Lambda_c \pi)=6.7^{+4.1}_{-3.8}\times 10^{-3}$
and $\mathcal{B}(\Lambda_b\rightarrow \Lambda_c K)=5.0^{+4.0}_{-3.0}\times 10^{-4}$,
where the uncertainties arise from the baryon light-cone distribution amplitudes (LCDAs),
the heavy charm quark and charmed baryon masses, and the hard scales.
It is shown that the asymmetry parameters in the two decays are approximately equal to -1,
which is expected as in the heavy quark limit and the soft meson limit.
Our predictions are consistent with the recent experimental data within errors.
The obtained numerical results are also compared to those in the other theoretical approaches when they are available.
\end{abstract}



\maketitle

\section{Introduction}
Weak decays of beauty baryons
offer a promising place to
extract  the Cabibbo-Kobayashi-Maskawa (CKM) matrix elements,
test  the heavy quark effective theory (HQET)~\cite{Grozin:1992yq,479387},
and search  for the effects of physics beyond the Standard Model (SM)
in a complementary field to the $B$ meson decays.
$\Lambda_b$ is the ground state of beauty baryons so it can only decay via weak interactions.
Because of the high $b$-quark mass, it has rich decay channels,
and provides a very good place to study  the semileptonic and nonleptonic  weak decays of heavy hadrons.
Among those, the decays containing a singly charmed baryon $\Lambda_c^+$ in the final state,
mediated by $b\rightarrow c$ quark transitions,
are expected to have sizable decay rates  due to the large color-allowed tree operator contributions.

Up to now several
semileptonic and nonleptonic decays of $\Lambda_b\rightarrow \Lambda_c$ were measured~\cite{jhep082014143,jhep042014087,prl112202001,prd79032001}.
A measurement of the  branching ratio   of   decay $\Lambda_b^0\rightarrow \Lambda_c^+\pi^-$
was performed by LHCb~\cite{jhep082014143,jhep042014087}
following observation of this decay by the CDF experiment~\cite{prl98122002} at the Tevatron collider.
The current world average of its branching ratio given by Particle Data Group (PDG)~\cite{pdg2020} is $(4.9\pm 0.4)\times 10^{-3}$.
The most abundant control sample of $\Lambda_b^0\rightarrow \Lambda_c^+\pi^-$ with $\Lambda_c^+$ decaying to $pK^-\pi^+$ final states,
are usually used to optimize the event selection and study systematic effects.
Taking the decay $\Lambda_b^0\rightarrow \Lambda_c^+\pi^-$ as the normalization mode,
LHCb~\cite{prd89032001} measured the relative branching ratio of the Cabibbo-suppressed partner $\Lambda_b^0\rightarrow \Lambda_c^+K^-$,
which has been considered in various analyses as a background component~\cite{prl109172003},
with respect to the decay  $\Lambda_b^0\rightarrow \Lambda_c^+\pi^-$ with significances of greater than $10\sigma$.
Subsequently, their absolute branching ratios are measured by LHCb~\cite{jhep042014087}
using another normalization channel of $B^0\rightarrow K^0\pi^+\pi^-$.
Very recently,  the first observation of the semitauonic decay $\Lambda_b^0\rightarrow \Lambda_c^+\tau^-\bar {\nu}_{\tau}$
is reported with a significance of $6.1\sigma$ by LHCb~\cite{220103497}.
As beauty baryons are now being observed in significant numbers in the LHCb detector,
new and more precise data will be available in the near future.
It will be interesting and timely to study weak decays of singly bottom baryons to final states
involving singly charmed baryons.

From the theoretical aspects,
there exist many phenomenological methods based on the SM
to explore the semileptonic and nonleptonic decays of
$\Lambda_b\rightarrow \Lambda_c$~\cite{prd531457,prd562799,prd58014016,Mohanta:1998iu,plb437403,prd80094016,ijmpa271250016,
prd90114033,prd91074001,prd95113002,epjc76284,plb815136125,prd97074007}.
The possible contributions of new physics to the semitauonic  decay and
the ratio of semileptonic branching fractions $\mathcal{R}(\Lambda_c)$
were analyzed in detail in Refs.~\cite{prd99055008,prd93054003,prd91115003,prd100113004,jhep02183,cpc45013113,prd99075006,
jhep092019103,jhep062021118,jhep022017068,jhep082017131,prd99015015,ijmpa331850169,prd100035035,jhep122019082,prd101113004,prd100113007,220105537}.
A recent first measurement of $\mathcal{R}(\Lambda_c)$  by the LHCb collaboration~\cite{220103497},
is still in agreement with the SM predictions.
The charmful decays of  $\Lambda_b^0\rightarrow \Lambda_c^+\pi^-,\Lambda_c^+K^-$
deserve special attention 
because they are uncontaminated by the contributions from the penguin operators.
These processes provide a good probe to test the factorization hypothesis,
which has been extensively explored in the heavy flavored meson case~\cite{Bauer:1986bm}.
Theoretical studies of these decays  have been performed using the soft-collinear effective theory (SCET)~\cite{plb586337},
light-front quark model (LFQM)~\cite{prd77014020,prd99014023}, Bethe-Salpeter (BS) model~\cite{mpla132265},
the nonrelativistic quark model (NRQM)~\cite{prd562799}, the light-front approach (LFA)~\cite{cpc42093101},
and QCD factorization (QCDF)~\cite{jhep092016112}. 
These calculations in the literature are important to check if the results in the SM are consistent with the experimental measurements.

The PQCD factorization theorem~\cite{prl744388,prd555577}  have been applied to deal with the
 exclusive heavy baryon decays~\cite{prd59094014,prd61114002},
in which the baryonic transition form factors or the decay amplitudes are factorized into the
 convolution of hard scattering kernels with the universal hadronic wave functions.
 The former can be perturbatively calculated in a systematic way,
 while the latter, absorbing nonperturbative contributions,
  are not calculable but universal.
Large logarithmic corrections are organized to all orders by Sudakov resummation to ensure a consistent perturbative expansion.
After determining the nonperturbative wave functions,
PQCD factorization theorem possesses predictive power.
Both factorizable and nonfactorizable contributions can be evaluated in a self-consistent manner within this approach.
The application of the extension of the PQCD approach to the baryon decays has achieved a preliminary success.
So far, the semileptonic heavy baryon decays $\Lambda_b\rightarrow p l \bar{\nu}$~\cite{prd59094014}
and $\Lambda_b\rightarrow \Lambda_c l \bar{\nu}$~\cite{prd61114002,cjp39328},
 the radiative decay $\Lambda_b\rightarrow \Lambda \gamma$~\cite{prd74034026},
 and the nonleptonic  decays $\Lambda_b\rightarrow  \Lambda J/\psi$~\cite{prd65074030},
 $\Lambda_b\rightarrow p \pi,pK$~\cite{prd80034011}
 have been studied systematically in the PQCD approach.
 Very recently, the  $\Lambda_b \rightarrow p$ transition form factors has been reanalyzed in PQCD
 by including higher-twist LCDAs of a $\Lambda_b$ baryon and a proton~\cite{220204804}.

The heavy-to-heavy decays are more complicated than the heavy-to-light ones,
because they involve an additional heavy baryon mass scale,
which provides a test of the application of the PQCD formalism. 
The semileptonic decay $\Lambda_b\rightarrow \Lambda_c$ have been studied by employing the diquark picture
in the so-called hybrid scheme~\cite{prd75054017},
in which the form factors in the small momentum transfer region are calculated in PQCD, while large one in HQET.
It has been found that perturbative contributions to the $\Lambda_b\rightarrow \Lambda_c$ decays
become more important at the maximal recoil of the $\Lambda_c$ baryon with the velocity transfer about 1.4.
This observation indicates that PQCD is an appropriate tool for analysis of
two-body nonleptonic  $\Lambda_b$  baryon decays.
Furthermore,
it was pointed out in Ref.~\cite{prd61114002} that  PQCD could be applicable
to $\Lambda_b\rightarrow \Lambda_c$  decays for velocity transfer greater than 1.2.
The determined $\Lambda_b$ and $\Lambda_c$ baryon wave functions at the leading twist accuracy
from the experimental data on the semileptonic $\Lambda_b\rightarrow \Lambda_c l\bar{\nu}$ decay,
can  be employed to study nonleptonic $\Lambda_b$ baryon decays because of their universality.

In this work,
we will investigate the two-body baryonic decays
$\Lambda_b\rightarrow \Lambda_c \pi, \Lambda_c K$ in the PQCD approach
to the leading order in the strong coupling $\alpha_s$,
for which
large momentum must be transferred to the two spectator quarks through the hard gluons exchange
so that they can  form collinear objects in the final state.
The decay amplitudes are calculated through Feynman diagrams involving $W$-emission, $W$-exchange, as well as three-gluons vertex diagrams.
Apart from the branching ratios, we also predict the parity-violating asymmetry parameter,
 which is related to the anisotropic angular distribution of the baryons produced in polarized baryon decays.
 The obtained results may provide valuable information to theoretical and experimental studies.

This paper is organized as follows: after the Introduction,
in Sec~\ref{sec:framework} we define the kinematic variables for the concerned $\Lambda_b$ decays
and the LCDAs for the associated hadrons.
Numerical results are presented and discussed in Sec~\ref{sec:results}.
Section~\ref{sec:sum} is devoted to our conclusion and outlook.
The explicit PQCD factorization formulas for the $T$-type decay amplitudes are collected in the Appendix.
\section{ FORMULATIONS}\label{sec:framework}
\begin{figure}[!htbh]
\begin{center}
\vspace{0.01cm} \centerline{\epsfxsize=8cm \epsffile{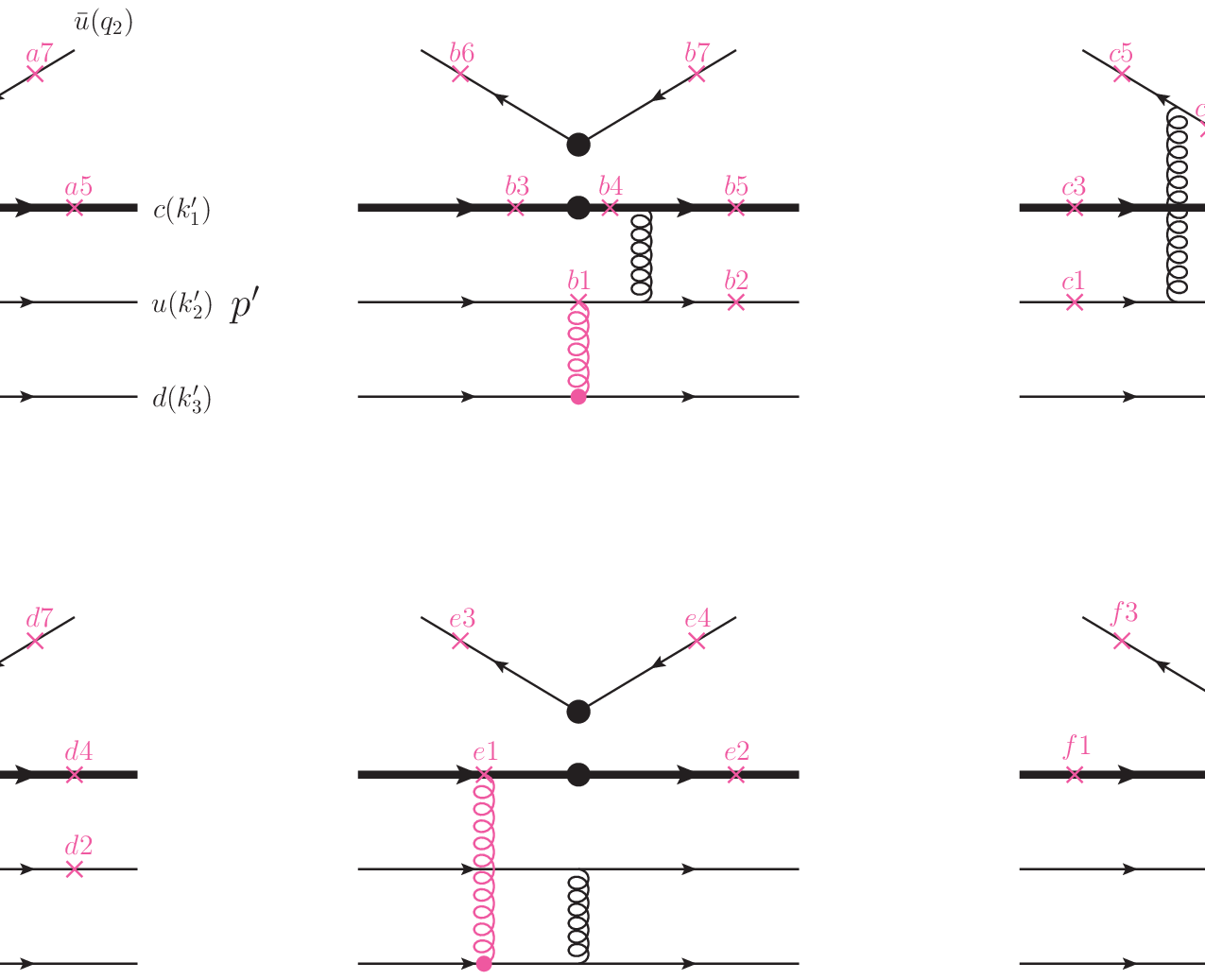}}
\setlength{\abovecaptionskip}{-2.0cm}
\caption{External $W$-emission ($T$) diagrams for the $\Lambda_b\rightarrow \Lambda_c \pi$  decay to the lowest order in the PQCD approach,
where the solid black blob represents the vertex of the effective weak interaction.
The heavy quarks are shown in bold lines.
The cross indicate the possible connections of the gluon shown in red attached to the $d$ quark. 
There are 36 diagrams in total, and we mark each one by $T_{ij}$ with $i=a-f$ and $j=1-7$.}
\label{fig:FeynmanT}
\end{center}
\end{figure}

\begin{figure}[!htbh]
\begin{center}
\vspace{0.01cm} \centerline{\epsfxsize=8cm \epsffile{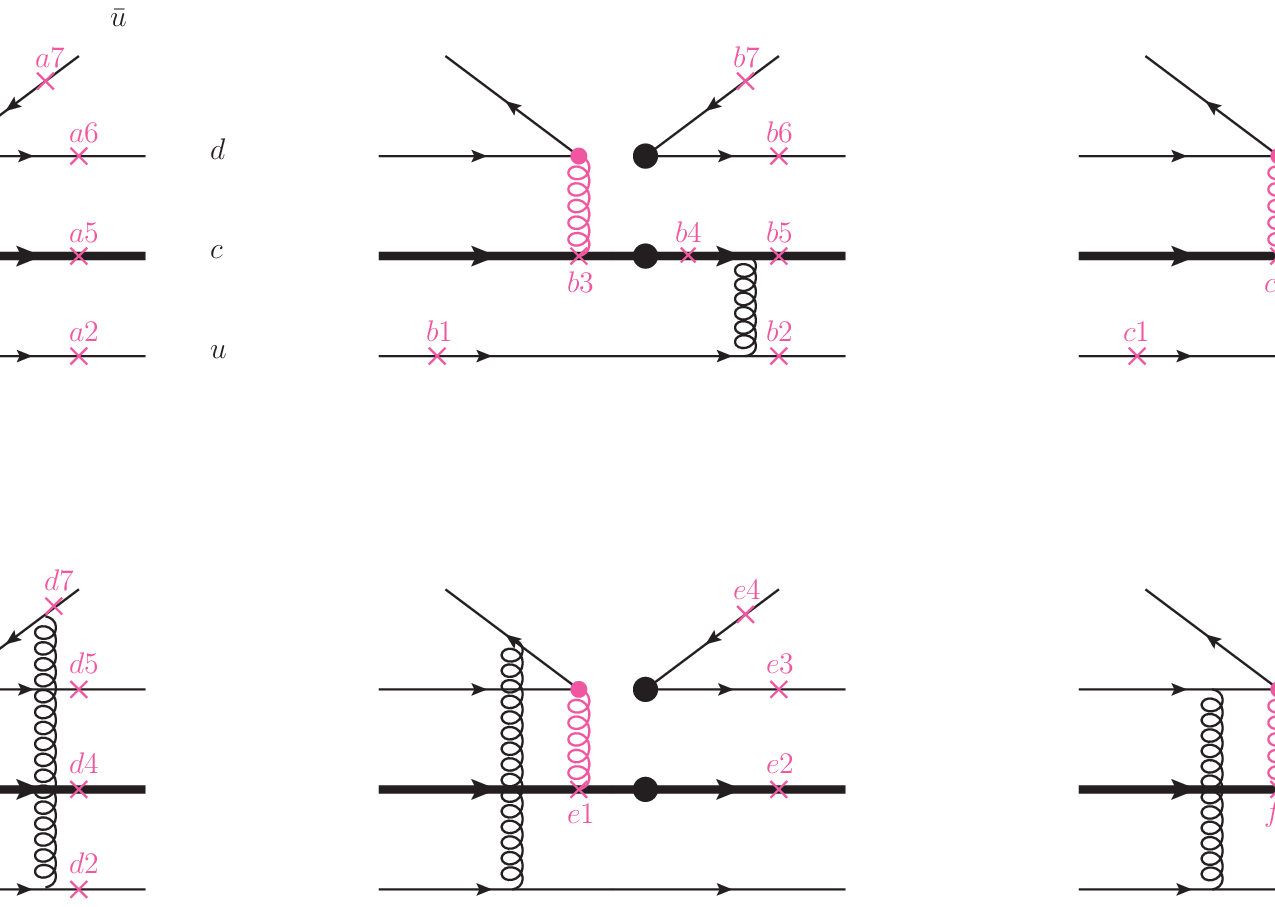}}
\setlength{\abovecaptionskip}{-3.0cm}
\caption{Internal $W$-emission ($C$) diagrams marked by $C_{ij}$ with $i=a-f$ and $j=1-7$.}
\label{fig:FeynmanC}
\end{center}
\end{figure}

\begin{figure}[!htbh]
\begin{center}
\vspace{0.01cm} \centerline{\epsfxsize=8cm \epsffile{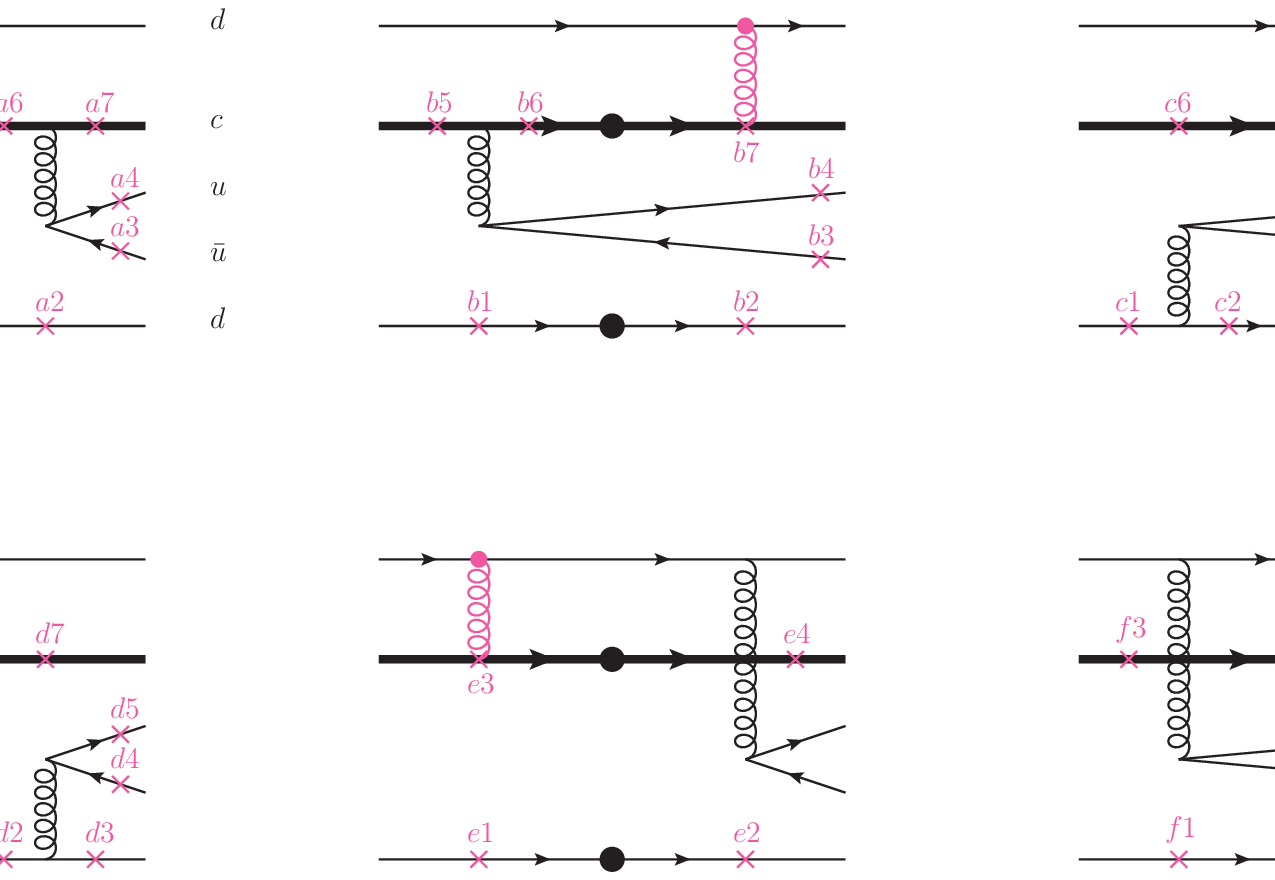}}
\setlength{\abovecaptionskip}{-3.0cm}
\caption{ $W$-exchange ($E$) diagrams marked by $E_{ij}$ with $i=a-f$ and $j=1-7$.}
\label{fig:FeynmanE}
\end{center}
\end{figure}

\begin{figure}[!htbh]
\begin{center}
\vspace{0.01cm} \centerline{\epsfxsize=8cm \epsffile{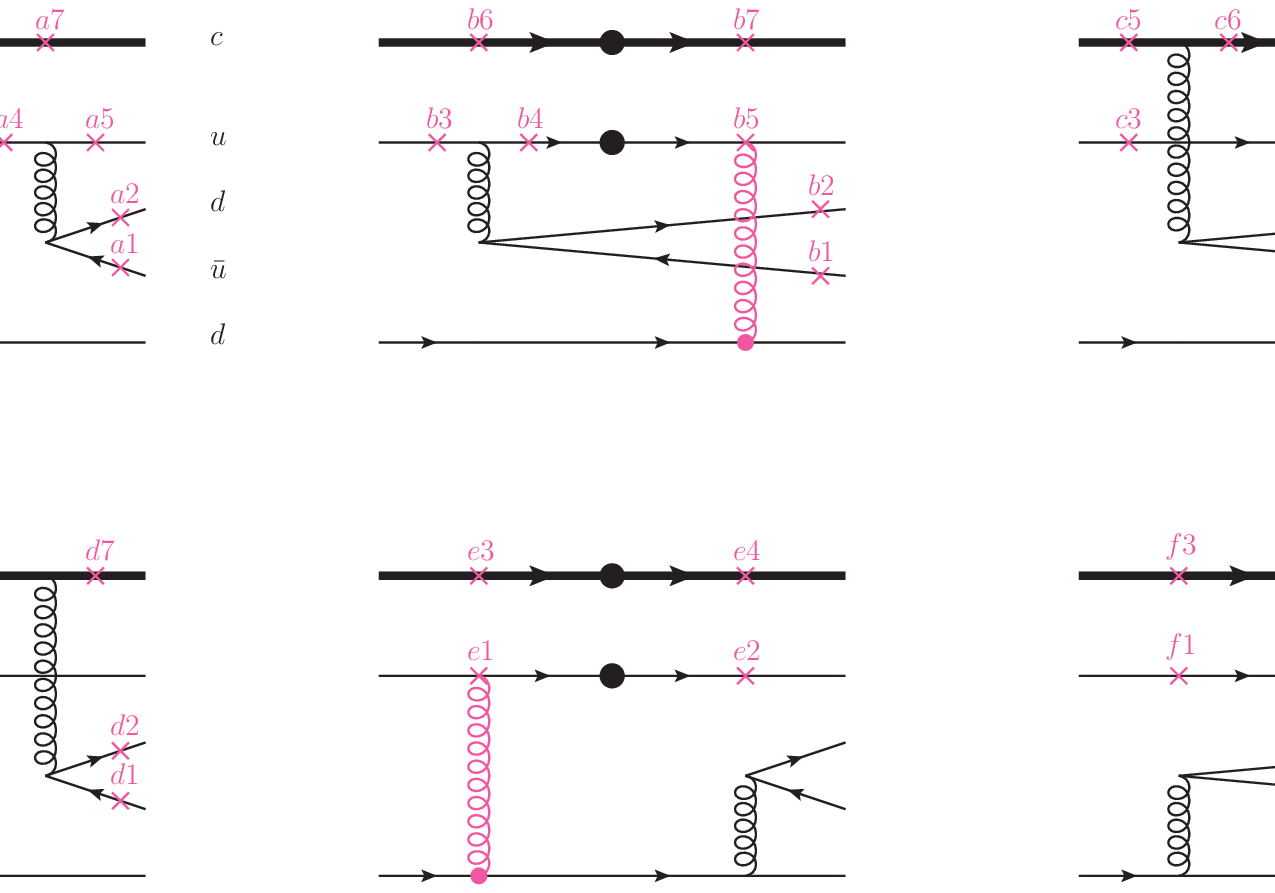}}
\setlength{\abovecaptionskip}{-3.0cm}
\caption{ $W$-exchange ($B$) diagrams marked by $B_{ij}$ with $i=a-f$ and $j=1-7$.}
\label{fig:FeynmanB}
\end{center}
\end{figure}

\begin{figure}[!htbh]
\begin{center}
\vspace{0.01cm} \centerline{\epsfxsize=8cm \epsffile{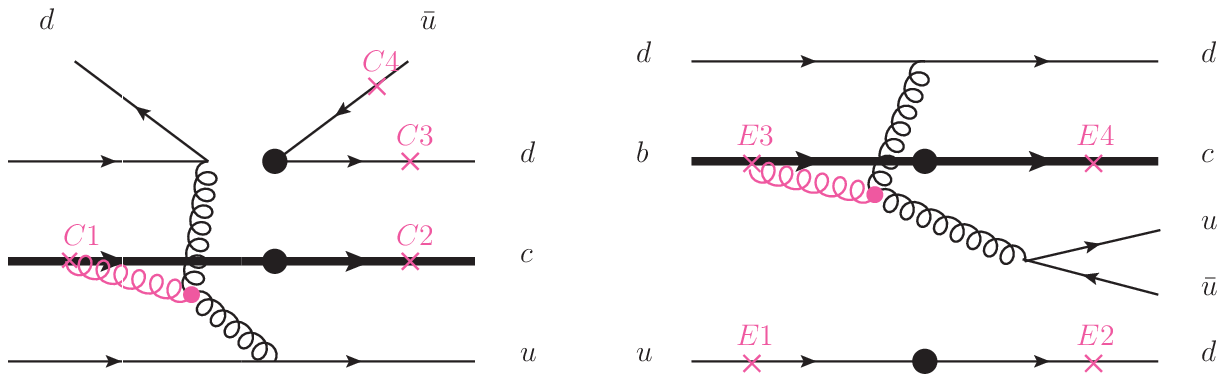}}
\setlength{\abovecaptionskip}{-6.0cm}
\caption{Three-gluons vertex ($G$) diagrams marked by $G_{ij}$ with $i=T,C,E,B$ and $j=1,2,3,4$.
The cross indicate the possible connections of the gluon shown in red attached to three-gluons vertex.}
\label{fig:FeynmanG}
\end{center}
\end{figure}

Because a baryon is composed of three constituent quarks in the conventional quark model, the QCD dynamics of baryon decay process  are more complicated than those of meson.
There are more possibilities for exchanging gluon, particularly in the case of the PQCD framework, which requires at least two hard gluon exchanges in the leading order approximation, as mentioned above.
For the nonleptonic two-body baryon decays, the hard amplitude involves eight external on shell quarks,
four of which correspond to the four-fermion operators and four of which are the spectator quarks in the initial and final states.
One must evaluate all possible Feynman diagrams for the eight-quark amplitude straightforwardly,
which include both factorizable and nonfactorizable contributions.
Taking $\Lambda_b\rightarrow \Lambda_c \pi$ as an example,
we can catalog these diagrams into five types: 
the external $W$-emission diagrams ($T$), the  internal $W$-emission diagrams ($C$),
the $W$-exchange diagram ($E$), the bow-tie contraction diagrams ($B$), and the three-gluons vertex diagrams ($G$),
as shown in Figs.~\ref{fig:FeynmanT},~\ref{fig:FeynmanC},~\ref{fig:FeynmanE},~\ref{fig:FeynmanB}, and~\ref{fig:FeynmanG}, respectively.

In the $T$ type diagrams, indicated by $T_{ij}$ in Fig.~\ref{fig:FeynmanT},
the two spectator quarks  in the initial state $\Lambda_b$
must be kicked by the hard gluons in order  to catch up the outgoing $c$ quark from the weak vertex
and form the color singlet, which further hadronize to a $\Lambda_c$ baryon.
The first subscript $i=a,b,c,d$ means one of the hard gluon connects the spectator $u$ quark to the four-quark operators,
which can be treated as a six-quark interaction like a meson decay in PQCD,
and $i=e,f$ denotes the gluon connects the two spectator quarks.
The second subscript $j$ represents possible ways of the other hard gluon connects the spectator $d$ quark with the six-quark system.
Excluding the case of two gluons
are simultaneously attached to one of the spectator quarks,
a total of 36 $T$-type diagrams contribute to the decay amplitude in the PQCD framework, as shown in Fig.~\ref{fig:FeynmanT}.

By exchanging the two identical $d$ quarks in the final states baryon and meson from the $T$-type diagrams,
one can obtain the 36 $C$-type diagrams, denoted by $C_{ij}$ in Fig.~\ref{fig:FeynmanC}.
When one of the gluons produces a pair of $u\bar u$, this corresponds to the $W$-exchange diagrams,
which are further classified into the $E$-type with the spectator $d$ quark entering into the $\Lambda_c$ baryon,
and the $B$-type with the spectator $d$ quark entering into the $\pi$ meson.
Each type have 36 diagrams, indicated by $E_{ij}$ and $B_{ij}$,
 as illustrated in Figs.~\ref{fig:FeynmanE} and~\ref{fig:FeynmanB}, respectively.
In addition, the concerned decays are admitted by the three-gluon contributions,
which  can also be divided into above four categories  with a total of 16 diagrams as shown in Fig.~\ref{fig:FeynmanG}.
Therefore, there are $36\times4+16=160$ diagrams in total that need to be evaluated to the leading order.
For the $\Lambda_b\rightarrow\Lambda_c K$ mode, the $d$ quark from the weak vertex is replaced by the $s$ quark,
thus there is no contributions from the $C$ and $B$-type diagrams.

We work in the rest frame of the $\Lambda_b$ baryon  with the  $\Lambda_c$ moving off in the plus direction,
and define $p$, $p'$, and $q$ to be four momenta of the $\Lambda_b$, $\Lambda_c$, and meson, respectively,
which can be written in the light-cone coordinates as
 \begin{eqnarray}\label{eq:pq}
p=\frac{M}{\sqrt{2}}\left(1,1,\textbf{0}_{T}\right),\quad
p'=\frac{M}{\sqrt{2}}\left(r^2,1,\textbf{0}_{T}\right),\quad
q=\frac{M}{\sqrt{2}}\left(1-r^2,0,\textbf{0}_{T}\right),
\end{eqnarray}
with the mass ratio $r=m/M$ and $m(M)$ is the mass of the $\Lambda_c(\Lambda_b)$ baryon.
The valence quark momenta  inside the initial and final state hadrons,
as shown in Fig.~\ref{fig:FeynmanT}, are parametrized as
 \begin{eqnarray}
k_1&=&\left(x_1\frac{M}{\sqrt{2}},\frac{M}{\sqrt{2}},\textbf{k}_{1T}\right),\quad
k_2=\left(x_2\frac{M}{\sqrt{2}},0,\textbf{k}_{2T}\right),\quad
k_3=\left(x_3\frac{M}{\sqrt{2}},0,\textbf{k}_{3T}\right),\nonumber\\
k_1'&=&\left(\frac{M}{\sqrt{2}}r^2,\frac{M}{\sqrt{2}}x_1',\textbf{k}'_{1T}\right),\quad
k_2'=\left(0,\frac{M}{\sqrt{2}}x_2',\textbf{k}'_{2T}\right),\quad
k_3'=\left(0,\frac{M}{\sqrt{2}}x_3',\textbf{k}'_{3T}\right),\nonumber\\
q_1&=&\left(y(1-r^2)\frac{M}{\sqrt{2}},0,\textbf{q}_{T}\right),\quad
q_2=\left((1-y)(1-r^2)\frac{M}{\sqrt{2}},0,-\textbf{q}_{T}\right).
\end{eqnarray}
Here, the $b$ and $c$ quarks are considered to be massive and carry momenta $k_1$ and $k'_1$, respectively,
while the masses of all light quarks and meson are neglected.
$y,x_l,x_l'$ are the parton longitudinal momentum fractions constrained by $0<y,x_l,x_l'<1$
and $\textbf{q}_{T}, \textbf{k}_{lT}, \textbf{k}'_{lT}$ are the corresponding perpendicular components.
They obey the conservation laws
\begin{eqnarray}
x_1+x_2+x_3=1,\quad \textbf{k}_{1T}+\textbf{k}_{2T}+\textbf{k}_{3T}=0,
\end{eqnarray}
which for the primed quantities takes a similar form.

As mentioned before, the universal nonperturbative hadronic wave functions,
which determines how quark momenta are distributed inside the hadron,
serve  as fundamental inputs for the PQCD  description of exclusive two-body nonleptonic processes.
The study of $\Lambda_b$ baryon wave function has made great progress in the past decade~\cite{plb665197,jhep112013191,epjc732302,plb738334,jhep022016179,Ali:2012zza},
since the complete classification of three-quark LCDAs
of the $\Lambda_b$ baryon in the heavy quark limit have been constructed~\cite{plb665197}.
The $\Lambda_b$ baryon LCDAs up to twist-4 has been available currently~\cite{jhep112013191,jhep022016179}
and several asymptotic models 
have been proposed in Refs.~\cite{plb665197,jhep112013191,Ali:2012zza}.
However, the $\Lambda_c$ one still receive less attention to date,
whose LCDAs beyond the leading twist are not yet available.
We thereby only take into account
the leading-twist LCDAs of baryons  in the following analysis.

As both $\Lambda_b$ and $\Lambda_c$ baryons are consist of
a heavy quark   and two light quarks,
the heavy quark can factorize from the light degrees of freedom in the heavy quark mass limit.
Assuming that
the two heavy baryons are in the ground state and
the orbital and spin degrees of freedom of the light quark systems decouple,
the leading-twist baryon LCDA can be expressed as~\cite{prd531416,zpc51321},
\begin{eqnarray}\label{eq:wave}
(\Psi_{\Lambda_Q})_{\alpha\beta\gamma}&=&\frac{1}{2\sqrt{2}N_c}\int\prod_{l=2}^3\frac{dy_l^-dy_l}{(2\pi)^3}
e^{ik_l\cdot y_l}\epsilon^{ijk}\langle 0|T[Q^i_\alpha(0)u_\beta^j(y_2)d^k_\gamma(y_3)]|\Lambda_Q(p)\rangle
\nonumber\\
&=&\frac{f_{\Lambda_Q}}{8\sqrt{2}N_c}[(\rlap{/}{p}+m_Q)\gamma_5 C]_{\beta\gamma}[\Lambda_Q(p)]_\alpha
\Phi_{\Lambda_Q}(k_i,\mu),
\end{eqnarray}
where $Q$ denotes the heavy quark $b$ or $c$, and $\Lambda_Q(p)(m_Q)$ is the corresponding heavy baryon spinor (mass).
$i,j,k$ and $\alpha,\beta,\gamma$ are the color and spinor indices, respectively.
$N_c$ is the number of colors. $C$ is the charge conjugation operator.
$f_{\Lambda_Q}$ is the normalization constant,
which satisfies the approximate relation $f_{\Lambda_b}m_{\Lambda_b}=f_{\Lambda_c}m_{\Lambda_c}$~\cite{prd61114002} in the heavy quark limit.
Neglecting the transverse momentum dependent,
the model for the $\Phi_{\Lambda_Q}$ employed in this work is taken from Ref.~\cite{9211255}
\begin{eqnarray}\label{eq:wave1}
\Phi_{\Lambda_Q}(x_1,x_2,x_3)=N_{\Lambda_Q}x_1x_2x_3\exp[-\frac{1}{2\beta_Q^2}(\frac{m_Q^2}{x_1}+\frac{m_q^2}{x_2}+\frac{m_q^2}{x_3})],
\end{eqnarray}
with $\beta_Q$ being the shape parameter
and $m_q$ the mass of light degrees of freedom in the baryon,
their numbers were determined to be $\beta_Q=1.0$ GeV and $m_q=0.3$ GeV in Ref.~\cite{prd61114002}
by fitting from the experimental data on the semileptonic decay $\Lambda_b\rightarrow \Lambda_c l\nu_l $ as mentioned before.
The normalization constants $N_{\Lambda_Q}$ are determined by
\begin{eqnarray}
\int dx_1dx_2dx_3\delta(x_1+x_2+x_3-1)\Phi_{\Lambda_Q}(x_1,x_2,x_3)=1.
\end{eqnarray}
Here we assume that the two baryon LCDAs possess  the same functional form and parameters except for different masses
since we know very little about the LCDAs of $ \Lambda_c$ baryon.
It is easy to observe that the LCDAs of heavy baryons are symmetric under permutation of two light quarks.
Above forms are supported by several phenomenological applications~\cite{prd61114002,prd75034011,prd59094014,prd65074030}.

The light-cone distribution amplitudes for the pseudoscalar mesons are given by~\cite{prd80034011} 
\begin{eqnarray}
\langle P(q)|\bar q_{2\beta}(z)q_{1\alpha}(0)|0\rangle=\frac{-i}{\sqrt{6}}\int_0^1dye^{iyq\cdot z}\gamma_5[\rlap{/}{q}\phi^A(y)
+m_0\phi^P(y)+m_0(\rlap{/}{n}\rlap{/}{v}-1)\phi^T(y)]_{\alpha\beta},
\end{eqnarray}
where $m_0$ is the chiral scale parameter of the pseudoscalar meson.
We use $m_0=1.4$ GeV and $m_0=1.9$ GeV for pion and kaon, respectively.
The kaon and pion meson distribution amplitudes up to twist-3 are determined using the light-cone QCD sum rules \cite{jhep01010,npb529323}:
\begin{eqnarray}
\phi_K^A(x)&=&\frac{3f_K}{\sqrt{6}}x(1-x)[1+a_1^KC_1^{3/2}(2x-1)+a_2^KC_2^{3/2}(2x-1)],\nonumber\\
\phi_K^P(x)&=&\frac{f_K}{2\sqrt{6}}[1+0.24C_2^{1/2}(2x-1)],\nonumber\\
\phi_K^T(x)&=&\frac{f_K}{2\sqrt{6}}(1-2x)[1+0.35(10x^2-10x+1)],\nonumber\\
\phi_{\pi}^A(x)&=&\frac{3f_{\pi}}{\sqrt{6}}x(1-x)[1+a_2^{\pi}C_2^{3/2}(2x-1)],\nonumber\\
\phi_{\pi}^P(x)&=&\frac{f_{\pi}}{2\sqrt{6}}[1+0.43C_2^{1/2}(2x-1)],\nonumber\\
\phi_{\pi}^T(x)&=&\frac{f_{\pi}}{2\sqrt{6}}(1-2x)[1+0.55(10x^2-10x+1)],\nonumber\\
\end{eqnarray}
with the Gegenbauer polynomials
\begin{eqnarray}
C_1^{3/2}(x)=3x,\quad C_2^{3/2}(x)=1.5(5x^2-1),\quad C_2^{1/2}(x)=(3x^2-1)/2.
\end{eqnarray}
The Gegenbauer moments for the twist-2 LCDAs are used with the following updated values at the scale $\mu=1$ GeV \cite{jhep05004}:
\begin{eqnarray}\label{eq:a1}
a_1^K=0.06\pm 0.03, \quad a_2^{K/\pi}=0.25\pm 0.15.
\end{eqnarray}

The effective Hamiltonian  
describing the decays under consideration is given by~\cite{Buchalla:1995vs}
\begin{eqnarray}
\mathcal{H}_{eff}&=&\frac{G_F}{\sqrt{2}} V_{cb}V^*_{uq}[C_1(\mu)O_1(\mu)+C_2(\mu)O_2(\mu)]+H.c.,
\end{eqnarray}
where the four-quark operators read as
\begin{eqnarray}
O_1(\mu)&=&(\bar c O_{\mu} u)( \bar q O^{\mu} b), \quad
O_2(\mu)=(\bar c O_{\mu} b)( \bar q O^{\mu} u),
\end{eqnarray}
with $O_{\mu}=\gamma_{\mu}(1-\gamma_5)$.
 $G_F$ is the Fermi coupling constant,
 $V_{cb},V_{uq}$ represent the CKM matrix elements with the quark $q=d(s)$ corresponding to $\pi(K)$ mode,
and $C_{1,2}(\mu)$ are the Wilson coefficients at the renormalization scale $\mu$.
Since the four quarks in the tree operators are different from each other,
the direct $CP$ asymmetries are absent naturally.

At the hadron level, the spin-dependent amplitude for decays of $\Lambda_b$ into $\Lambda_c$ and a light pseudoscalar meson $P$
is obtained by sandwiching the effective Hamiltonian between the initial and final hadron states,
\begin{eqnarray}\label{eq:ab}
\mathcal{M}&=&\langle \Lambda_c P|H_{eff}|\Lambda_b\rangle =\bar \Lambda_c(p')[M_S+M_P\gamma_5]\Lambda_b(p),
\end{eqnarray}
where we split the amplitudes into
the $S$-wave ($M_S$) and $P$-wave ($M_P$) pieces,
corresponding to the parity violating and conserving ones.
Their generic factorization formula
can be written as
\begin{eqnarray}
M_{S(P)}=\frac{f_{\Lambda_b}f_{\Lambda_c}\pi^2 V_{cb}V^*_{uq} G_F}{144\sqrt{3}}\sum_{R_{ij}} \int [\mathcal{D}x][\mathcal{D}b]_{R_{ij}}
\alpha_s^2(t_{R_{ij}})a(t_{R_{ij}})\Phi_{\Lambda_b}(x)\Phi_{\Lambda_c}(x')H^{S(P)}_{R_{ij}}(x,x',y)\Omega_{R_{ij}}(b,b',b_q) e^{-S_{R_{ij}}},
\end{eqnarray}
where $R=T,C,B,E,G$ labels the five type diagrams as discussed above.
The summation extends over all possible diagrams.
$a$ is an appropriate combination of the Wilson coefficients.
$H_{R_{ij}}$ is the numerator of the hard amplitude depending on the spin structure of final state,
while $\Omega_{R_{ij}}$ is the Fourier transformation of the denominator of the hard amplitude from the $k_T$ space to its conjugate $b$ space.
These quantities associated with specific diagram are collected in Appendix.
The integration measure of the momentum fractions  can be written as
\begin{eqnarray}
[\mathcal{D}x]=[dx][dx']dy, \quad [dx]=dx_1dx_2dx_3\delta(1-x_1-x_2-x_3),\quad [dx']=dx'_1dx'_2dx'_3\delta(1-x'_1-x'_2-x'_3),
\end{eqnarray}
and the measure of the transverse extents $[\mathcal{D}b]$ are also given   in the Appendix.

The explicit form of the Sudakov factors appearing in the above equations are given by~\cite{prd80034011}
\begin{eqnarray}
S_{R_{ij}}=\sum_{l=2,3}s(w,k^+_l)+ \sum_{l=1,2,3}s(w',k'^-_l)+
\frac{8}{3}\int^{t_{R_{ij}}}_{\kappa w}\frac{d\bar \mu}{\bar \mu}\gamma_q(\alpha_s(\bar \mu))+3\int^{t_{R_{ij}}}_{\kappa w'}\frac{d\bar \mu}{\bar \mu}\gamma_q(\alpha_s(\bar \mu)),
\end{eqnarray}
\begin{eqnarray}
S_{R_{ij}}&=&\sum_{l=2,3}s(w,k^+_l)+ \sum_{l=1,2,3}s(w',k'^-_l)+ \sum_{l=1,2}s(w_q,q^+_l)+\nonumber\\&&
\frac{8}{3}\int^{t_{R_{ij}}}_{\kappa w}\frac{d\bar \mu}{\bar \mu}\gamma_q(\alpha_s(\bar \mu))
+3\int^{t_{R_{ij}}}_{\kappa w'}\frac{d\bar \mu}{\bar \mu}\gamma_q(\alpha_s(\bar \mu))
+2\int^{t_{R_{ij}}}_{\kappa w_q}\frac{d\bar \mu}{\bar \mu}\gamma_q(\alpha_s(\bar \mu)),
\end{eqnarray}
for the factorizable and nonfactorizable diagrams, respectively.
The explicit expression of the function $s$ can be found in Ref.~\cite{npb32562}.
Another threshold Sudakov factor $S_t(x)$, collected the double logarithms $\alpha_s \ln^2(x)$ to all orders,
is set to 1 similar to the case of the heavy-to-light $\Lambda_b$ decays~\cite{220204804}.

The hard scale $t$ for each diagram should be chosen as the maximal virtuality of internal particles in a hard amplitude,
\begin{eqnarray}
t_{R_{ij}}=\max(\sqrt{|t_A|},\sqrt{|t_B|},\sqrt{|t_C|},\sqrt{|t_D|},w,w',w_q),
\end{eqnarray}
where the hard scales $t_{A,B,C,D}$ associated with the two hard gluons and two virtual quark,
whose expressions will be listed in Appendix.
The factorization scales $w$, $w'$, and $w_q$ are taken to be
 \begin{eqnarray}
w^{(')}=\min(\frac{1}{b^{(')}_1},\frac{1}{b^{(')}_2},\frac{1}{b^{(')}_3}),\quad w_q=\frac{1}{b_q},
\end{eqnarray}
 with the variables
 \begin{eqnarray}
b^{(')}_1=|b^{(')}_2-b^{(')}_3|,
\end{eqnarray}
with the other $b^{(')}_l$ defined by permutation.
The phenomenological factor $\kappa=1.14$ is adopted according to Ref.~\cite{epjc8637}.

\section{Numerical results}\label{sec:results}
In this section, we first present the input parameters  entering our numerical analysis.
The SM parameters such as baryon and heavy quark masses (GeV), lifetimes (ps), and the Wolfenstein parameters for the CKM matrix
are taken from PDG~\cite{pdg2020},
\begin{eqnarray}
M&=&5.6196 \quad  m=2.286  \quad m_b=4.8, \quad m_c=1.275,\quad \tau=1.464,\nonumber\\
\lambda &=& 0.22650, \quad  A=0.790,  \quad \bar{\rho}=0.141, \quad \bar{\eta}=0.357.
\end{eqnarray}
Other parameters appearing in the baryon and meson DAs have been specified before.

We first examine the baryonic transition form factors at maximal recoil,
to which the factorizable emission diagrams from Fig.~\ref{fig:FeynmanT} are related.
The $\Lambda_b\rightarrow \Lambda_c$ transition matrix elements induced by the vector and axial-vector currents 
can each be decomposed into three dimensionless invariant form factors~\cite{prd91074001,prd90114033}
\begin{eqnarray}\label{eq:form}
\langle  \Lambda_c(p')|\bar c \gamma^\mu b|\Lambda_b(p)\rangle &=&\bar{\Lambda}_c(p')[f_1(q^2)\gamma^\mu
-\frac{f_2(q^2)}{M}i\sigma^{\mu\nu}q_\nu+\frac{f_3(q^2)}{M}q^\mu]\Lambda_b(p),\nonumber\\
\langle  \Lambda_c(p')|\bar c \gamma^\mu\gamma_5 b|\Lambda_b(p)\rangle &=&\bar{\Lambda}_c(p')[g_1(q^2)\gamma^\mu
-\frac{g_2(q^2)}{M}i\sigma^{\mu\nu}q_\nu+\frac{g_3(q^2)}{M}q^\mu]\gamma_5\Lambda_b(p),
\end{eqnarray}
where $\sigma^{\mu\nu}=i(\gamma^\mu\gamma^\nu-\gamma^\nu\gamma^\mu)/2$ and  all $\gamma$ matrices are defined as in Bjorken-Drell.
$q=p-p'$ is the four-momentum transfer constrained by the physical kinematic region $0<q^2<(M-m)^2$.
 Contracting Eq.~(\ref{eq:form}) with $q_\mu$, we have
 \begin{eqnarray}\label{eq:form1}
\langle  \Lambda_c(p')|\bar c \rlap{/}{q}(1-\gamma_5) b|\Lambda_b(p)\rangle &=&\bar{\Lambda}_c(p')[f_1(q^2)\rlap{/}{q}
-g_1(q^2)\rlap{/}{q}\gamma_5]\Lambda_b(p)\nonumber\\&=&\bar{\Lambda}_c(p')[f_1(q^2)(M-m)
+g_1(q^2)(M+m)\gamma_5]\Lambda_b(p),
\end{eqnarray}
in which the free Dirac equation  has been used.
Here, $f_2$ and $g_2$ terms vanish due to the antisymmetric structure of $\sigma^{\mu\nu}$,
while the contributions from $f_3$ and $g_3$  are neglected in the massless limit $q^2\rightarrow 0$.
We shall concentrate on $f_1$ and $g_1$ in the present work.
It can be seen the right-hand side of above equation has a structure similar to
the corresponding one for the decay amplitude  in Eq.~(\ref{eq:ab}).
Thus the baryonic transition form factors  $f_1$ and $g_1$  at the zero momentum transfer 
can be extracted from the decay amplitudes
through the relations,
\begin{eqnarray}
M_S^{\text{fac}}&=&\frac{G_F}{\sqrt{2}} V_{cb}V^*_{uq}(M-m)f_P(C_2+\frac{1}{3}C_1)f_1(0),\nonumber\\
M_P^{\text{fac}}&=&\frac{G_F}{\sqrt{2}} V_{cb}V^*_{uq}(M+m)f_P(C_2+\frac{1}{3}C_1)g_1(0),
\end{eqnarray}
where $M_S^{\text{fac}}$ and $M_P^{\text{fac}}$ denotes the factorizable contributions to the corresponding amplitudes $M_S$ and $M_P$, respectively.
$f_P$ is the decay constant of the  pseudoscalar meson.
The diagrams that contribute to
the form factors from $T_{a1-a5}$, $T_{b1-b5}$, $T_{e1,e2}$, and $T_{f1,f2}$ in Fig.~\ref{fig:FeynmanT},
whereas those from the triple-gluon vertex, such as $G_{T1,T2}$ in Fig.~\ref{fig:FeynmanG},
do not contribute since their color factors are zero in the present case.

\begin{table}[!htbh]
\footnotesize
\caption{Theoretical predictions for the form factors $f_1$ and $g_1$ at $q^2=0$ of $\Lambda_b\rightarrow \Lambda_c$ transition
using different approaches.}
\label{tab:form}
\begin{tabular}[t]{lccccccccccc}
\hline\hline
Form factors & This work 
& \cite{prd94073008}&\cite{prd91074001}&\cite{prd92034503} &\cite{prd77014020}
&\cite{prd99054020}&\cite{cpc42093101} &\cite{prd100034025} &\cite{epjc79540}&\cite{prd104013005}&\cite{prd102034033}\\  \hline
$f_1(0)$ &$0.440_{-0.082-0.149-0.035}^{+0.113+0.014+0.074}$ &0.526 &0.549&$0.418\pm0.161$  &0.506 &$0.500^{+0.028}_{-0.031}$
& 0.670 &$0.474^{+0.069}_{-0.072}$ &0.488&$0.50\pm0.05$ &$0.50\pm0.00$\\
$g_1(0)$ &$0.443_{-0.085-0.156-0.035}^{+0.120+0.013+0.076}$ &0.505 & 0.542&$0.378\pm0.102$  &0.501 &$0.509^{+0.029}_{-0.031}$
& 0.656 &$0.468^{+0.067}_{-0.070}$ &0.470&$0.49\pm0.05$ &$0.50\pm0.00$\\
\hline\hline
\end{tabular}
\end{table}
The numerical results of the form factors are presented in Table~\ref{tab:form},
where those from various theoretical calculations in the literature are also shown for comparison.
The main theoretical uncertainties stem from
the shape parameters $\beta_Q$ in the baryon LCDAs,
the charm quark and charmed baryon masses, and the hard scales $t$, respectively.
In the evaluation, we vary the values of $\beta_Q$, $m_c$ and $m$ within a $10\%$ range
and the hard scale from $0.75t$ to $1.25t$ for the error estimation.
It is found that the main uncertainties in our calculations come from the baryon LCDAs,
which can reach $25\%$ in magnitude.
It is necessary to stress that the higher-twists contributions of the baryon LCDAs are neglected
because that of the $\Lambda_c$ one has not been available currently,
which may give significant uncertainties.
So we choose a relatively wide range of $\beta_Q$  to estimate the effects of the higher-twist contributions.
The large uncertainties imply that the nonperturbative parameters in the DAs of baryons
need to be further constrained for  improving the precision of theoretical predictions.
The uncertainties stemming from the baryon decay constant $f_{\Lambda_b}$ are not shown explicitly in the Table,
whose effect on the form factors via the relation of $f_0(g_0) \propto f_{\Lambda_b}^2$.
The uncertainties related to the light pseudoscalar mesons,
such as the Gegenbauer moments shown in Eq.~(\ref{eq:a1}), are only several percent,
so that they can be safely neglected in our analysis.

One can see that the two form factors are nearly equal as expected in the heavy quark limit.
Our results are comparable with those calculated from other approaches.
Most of the theoretical model calculations predict comparable values for the form factors
ranging from 0.4  to 0.7.
The main difference in various predictions from the LFQM are in the treatment of the light quarks inside baryons.
The baryon quark-diquark picture is widespread adopted in Refs.~\cite{prd77014020,prd99054020,cpc42093101,prd100034025},
in which the two light spectator quarks in baryon is considered as a scalar of color antitriplet.
Instead, the authors of Ref.~\cite{epjc79540}
treat each of the three constituent quarks in baryon as separate dynamic entities.
It was shown that  the quark-diquark picture works well for heavy baryons.
Reference~\cite{prd104013005} worked in the triquark scheme,
in which the three-body wave functions are based on the baryon spectroscopy.
In addition, the relativistic quark model (RQM)~\cite{prd94073008}, the covariant confined quark model (CCQM)~\cite{prd91074001},
and the lattice QCD (LQCD)~\cite{prd92034503} also give similar predictions.
In Ref.~\cite{prd75054017},
 the so-called hybrid scheme is employed,
in which the form factor in the region with large $q^2$ is derived in the framework of the HQET,
while the small region is evaluated utilizing the conventional PQCD formalism,
which is free from the endpoint singularity.
Their  form factor $f_1$(or $g_1$\footnote{Note that the definition of $g_1$ in~\cite{prd75054017} differs from ours in sign.})
 was parametrized  in the so-called ``Isgur-Wise function," 
\begin{eqnarray}\label{eq:iwf}
f_1(\rho)=|g_1(\rho)|=1-3.61(\rho-1)+7.24(\rho-1)^2-5.83(\rho-1)^3,
\end{eqnarray}
with $\rho=\frac{p\cdot p'}{Mm}$.
One can obtain the value of $f_1$ at $q^2=0$, corresponding to $\rho=\frac{M^2+m^2}{2Mm}$, is 0.32,
which is somewhat small.
We note that the magnitude of the slope in Eq.~(\ref{eq:iwf})
is obviously greater than the LHCb measurement of $1.63\pm0.07\pm0.08$~\cite{prd96112005}.
As stated in Ref.~\cite{prd75054017}, the slope of the Isgur-Wise function is model dependent,
the high power terms would compensate for the deviation of the linear terms.
For more widely discussed about the slope, see the recent analysis in Ref.~\cite{epjc80926} and references therein.
In an earlier work~\cite{prd61114002}, the authors
have developed a PQCD factorization theorem for the semileptonic heavy baryon decay $\Lambda_b\rightarrow \Lambda_c l \nu$,
where the form factors are parametrized by $f_1(\rho)=\frac{1.32}{\rho^{5.18}}$ and  $g_1(\rho)=-\frac{1.19}{\rho^{5.14}}$.
This allows us to evaluate the form factors to be $f_1=0.20$ and $|g_1|=0.18$ at $q^2=0$,
which are half of ours.
The reason is that they neglected the transverse momentum dependence of the internal quark propagators,
and assumed that  charm quark and $\Lambda_c$ baryon have the equal mass, namely $m_c\approx m$.
However, the mass difference, $m-m_c$, can reach as larger as 1 GeV, which is really important numerically.
As we will see later, keeping the nonzero mass difference makes the branching ratios of the corresponding two-body decay
more coincides with the data.

The decay branching ratio and up-down asymmetry of the concerned decays
for the initial baryon in the rest frame are given as~\cite{prd423746,prd562799}
\begin{eqnarray}
\mathcal{B}&=&\frac{|P|\tau_{\Lambda_b}}{8\pi}[(1+r)^2|M_S|^2+(1-r)^2|M_P|^2], \nonumber\\
\alpha &=&-\frac{2(1-r^2) Re[M_S^*M_P]}{(1+r)^2|M_S|^2+(1-r)^2|M_P|^2},
\end{eqnarray}
where $|P|$ is the magnitude of the three momentum
of the $\Lambda_c$ baryon in the rest frame of the $\Lambda_b$ baryon.

We first investigate the relative importance of the different topologies
contributions to the decay amplitudes,
whose results are displayed separately in Table~\ref{tab:amplitude}.
It is found  that both the $S$-wave and $P$-wave decay amplitudes
are governed by the $T$-type diagrams as it should be,
while others are at least one order smaller.
Our numerical results show the contributions of the $T$-type diagrams accounts for more than $90\%$ of the total branching ratio,
even reaching $98\%$ for the kaon mode due to the vanishing $C$-and $B$-types amplitudes as mentioned above.
The contribution from the three-gluons vertex diagrams is found to be less important compared with the $W$-emission diagrams
but greater than the $W$-exchange ones.
The observed hierarchy pattern for their relative contribution satisfy $T>C>G>E>B$,
which seems compatible with the previous predictions in Ref.~\cite{plb586337},
although the contributions from the three-gluons vertex are not yet considered in Ref.~\cite{plb586337}.

As noticed before, the factorizable diagrams contain  14 diagrams of $T$-type and 2 diagrams of $G$-type,
while the remaining ones are all classified as nonfactorizable contributions.
In Table~\ref{tab:fac}, we present the factorizable and nonfactorizable contributions in the decay amplitudes.
It is observed that the
factorizable contributions dominate over the nonfactorizable ones.
This situation differs from the PQCD predictions on the decays of
$\Lambda_b\rightarrow p \pi$~\cite{prd80034011}, 
where the nonfactorizable contributions are  found almost 2 orders of
magnitude larger than those from the factorizable ones.
As stated in Ref.~\cite{prd80034011}, in the $k_T$ factorization approach,
the Sudakov factor can only suppress the small transverse momentum region,
and has almost no effect in the large region.
In some nonfactorizable diagrams, the two virtual quarks can
be on the mass shell in the large transverse momentum region,
so that their contributions are not subjected to the suppression from the Sudakov factor.
We find a similar situation also exist in some of the factorizable diagrams
of the heavy-to-heavy baryonic decays, in which  the heavy charm quark and charmed baryon mass can not be negligible.
Taking the factorizable diagram $T_{b4}$ in Fig.~\ref{fig:FeynmanT} for example,
as can be seen in Table~\ref{tab:wil}, the masses of charm quark and charmed baryon
enters into the denominator of the propagators of the virtual quarks,
and makes the two virtual quarks are on the mass shell in the large transverse momentum region.
It can numerically change the real and imaginary parts of the corresponding amplitude and enhance its module.
The numerical analysis  shows that the two factorizable
diagrams $T_{b4}$ and $T_{b5}$ play the most significant role in the decays under study.
The authors in Ref.~\cite{prd65074030} also found that the dominant nonfactorizable contributions in $\Lambda_b\rightarrow \Lambda  J/\psi$ decay.
It is understandable that this channel is a color-suppressed type as the $B\rightarrow J/\psi K$ decay,
where the factorizable contributions are suppressed by the small Wilson coefficient.

\begin{table}
\caption{ The values of decay amplitudes from different type Feynman diagrams for $\Lambda_b\rightarrow \Lambda_c \pi(K)$  decays.}
\label{tab:amplitude}
\begin{tabular}[t]{lccc}
\hline\hline
Type &$M_S$ &$M_P$ \\ \hline
$\Lambda_b\rightarrow \Lambda_c \pi$ \\
T    & $-2.0\times 10^{-8}+i8.5\times 10^{-8}$ & $-4.6\times 10^{-8}+i2.0\times 10^{-7}$ \\ 
C    & $-2.0\times 10^{-9}+i4.4\times 10^{-9}$ & $-6.3\times 10^{-9}+i1.2\times 10^{-8}$ \\ 
E    & $-6.5\times 10^{-10}+i1.4\times 10^{-9}$ & $-1.1\times 10^{-9}-i1.6\times 10^{-10}$\\ 
B    & $1.8\times 10^{-10}-i9.1\times 10^{-10}$ & $5.7\times 10^{-10}-i2.0\times 10^{-9}$ \\ 
G    & $1.6\times 10^{-9}-i1.7\times 10^{-9}$ & $3.7\times 10^{-9}-i3.6\times 10^{-9}$  \\ 
$\Lambda_b\rightarrow \Lambda_c K$  \\
T    &$-5.6\times 10^{-9}+i2.4\times 10^{-8}$ &$-1.3\times 10^{-8}+i5.6\times 10^{-8}$ \\ 
E    &$-1.4\times 10^{-10}+i3.2\times 10^{-10}$ &$-2.7\times 10^{-10}-i8.4\times 10^{-11}$ \\ 
G    &$-7.2\times 10^{-11}-i4.0\times 10^{-11}$ &$-2.1\times 10^{-10}-i6.2\times 10^{-11}$ \\ 
\hline\hline
\end{tabular}
\end{table}

\begin{table}
\caption{ The values of decay amplitudes from the factorizable and nonfactorizable diagrams for $\Lambda_b\rightarrow \Lambda_c \pi(K)$  decays.}
\label{tab:fac}
\begin{tabular}[t]{lcc}
\hline\hline
Amplitude &Factorizable &Nonfactorizable  \\ \hline
$M_S(\Lambda_b\rightarrow \Lambda_c \pi)$  & $-1.7\times 10^{-8}+i7.4\times 10^{-8}$ & $-3.9\times 10^{-9}+i1.4\times 10^{-8}$ \\
$M_P(\Lambda_b\rightarrow \Lambda_c \pi)$  & $-4.1\times 10^{-8}+i1.8\times 10^{-7}$ & $-8.1\times 10^{-9}+i2.6\times 10^{-8}$ \\
$M_S(\Lambda_b\rightarrow \Lambda_c K)$    & $-4.6\times 10^{-9}+i2.0\times 10^{-8}$ & $-1.2\times 10^{-9}+i4.3\times 10^{-9}$\\
$M_P(\Lambda_b\rightarrow \Lambda_c K)$    & $-1.1\times 10^{-8}+i4.8\times 10^{-8}$ & $-2.5\times 10^{-9}+i7.9\times 10^{-9}$ \\
\hline\hline
\end{tabular}
\end{table}

\begin{table}[!htbh]
\tiny
\caption{Various theoretical results on the branching ratios ($10^{-3}$) and the up-down asymmetries ($\%$) of $\Lambda_b\rightarrow \Lambda_c \pi(K)$ decays.}
\label{tab:branching}
\begin{tabular}[t]{lccccccccccccc}
\hline\hline
Mode & This work &\cite{zpc59179} &\cite{prd562799}\footnotemark[1] &\cite{prd575632,mpla13181}\footnotemark[1] &\cite{mpla1323} &\cite{Mohanta:1998iu} &\cite{prd99054020}&\cite{epjc79540}\footnotemark[1] &\cite{prd100034025} &\cite{prd102034033} & \cite{prd99055008}
 & \cite{jhep092016112}&PDG2020~\cite{pdg2020}\\ \hline
$\mathcal{B}(\Lambda_b\rightarrow \Lambda_c \pi)$ &$6.7_{-2.2-2.8-1.3}^{+3.2+0.3+2.5}$ &$4.6^{+2.0}_{-3.1}$ &4.5 &5.6 &3.91 &1.75
&4.96 &3.8 &$4.16^{+2.43}_{-1.73}$ &$4.5\pm0.2$ &$3.6\pm0.3$ &$2.85\pm0.54$ &$4.9\pm0.4$ \\
$\mathcal{B}(\Lambda_b\rightarrow \Lambda_c K)$ &$0.5_{-0.2-0.2-0.1}^{+0.3+0.0+0.2}$ & $\cdots$& $\cdots$ & $\cdots$& $\cdots$& $0.13$
 & $0.393$ &0.31 &$0.31^{+0.18}_{-0.13}$  &$0.34\pm0.01$ & $\cdots$ &$0.221\pm0.040$ &$0.359\pm0.030$ \\
$\alpha(\Lambda_b\rightarrow \Lambda_c \pi)$ &-99.2 &-100 &-99 & -99 & $\cdots$ &-99.9 &-99.8 &-99.9 &$-99.99^{+4.70}_{-0.00}$
&-$100.0\pm0.0$ & $\cdots$& $\cdots$& $\cdots$ \\
$\alpha(\Lambda_b\rightarrow \Lambda_c K)$  &-98.2 & $\cdots$ & $\cdots$& $\cdots$ & $\cdots$
&-100 &-100 &-99.9&$-99.97^{+5.02}_{-0.01}$&-$100.0\pm0.0$ & $\cdots$ & $\cdots$& $\cdots$\\
\hline\hline
\end{tabular}
\footnotetext[1]{We estimate the branching ratio by multiplying the given width by the lifetime $\tau_{\Lambda_b}=1.464$ps.}
\end{table}

The predicted branching ratios and up-down asymmetries  are summarized in Table~\ref{tab:branching},
together with the results of other theoretical studies and the experimental data.
The sources of theoretical errors are the same as in Table~\ref{tab:form}.
A variety of predictions derived from the quark models based on  factorization assumption
have been conducted,
with wide-ranging predictions for $\mathcal{B}(\Lambda_b\rightarrow \Lambda_c \pi)\sim (1.75-5.6)\times 10^{-3}$
and $\mathcal{B}(\Lambda_b\rightarrow \Lambda_c K)\sim (1.3-3.93)\times 10^{-4}$.
Overall speaking, our results are slightly larger than other theoretical predictions due to the involved important
nonfactorizable effects.
In the absence of the nonfactorizable contributions, our branching ratios for the pion and kaon modes  will be reduced to
$4.9\times 10^{-3}$ and $3.5\times 10^{-4}$, respectively,
which seem  to be more consistent with those predictions based on the factorization assumption as shown in Table~\ref{tab:branching}.
It is understandable because the naive factorization suffices to describe the color-allowed type decays.
From Table~\ref{tab:fac}, one can see that a constructive interference between the factorizable contributions
and the nonfactorizable ones enhance the branching ratios by about $30\%$,
indicate they are indeed numerically significant in the considered processes.
The important nonfactorizable contributions were also found in previous studies based on the relativistic three-quark model~\cite{prd575632,mpla13181}, but with the destructive interference pattern.
Previous measurements performed by the LHCb collaboration~\cite{jhep042014087} yield
$\mathcal{B}(\Lambda_b\rightarrow \Lambda_c \pi)\sim (5.97\pm0.86)\times 10^{-3}$ and
$\mathcal{B}(\Lambda_b\rightarrow \Lambda_c K)\sim (3.55\pm0.66)\times 10^{-4}$, respectively,
where the statistical and systematic uncertainties are combined in quadrature.
Our prediction on the pion mode is consistent with its value,
while that of kaon mode is larger.
Although the inclusion of the nonfactorizable contribution  makes our results  larger,
for  a consistent and complete analysis of  the leading-order PQCD calculation,
it is necessary to include the nonfactorizable contribution in the current work.

In the heavy quark limit and the soft meson limit,
one expect that $M_S\approx M_P(1-r)/(1+r)$, resulting in $\alpha\approx -1$.
As given in the Table~\ref{tab:branching},
almost all the predicted asymmetry parameters are nearly $100\%$ and negative,
which indicate the $V-A$ nature of the weak current and maximum parity violation.
The values of $\alpha$ are insensitive to the variation of these parameters,
indicates  it can serve as an ideal quantity to test the PQCD approach.

The ratio of kaon to pion decay rates is defined by
\begin{eqnarray}\label{eq:rr}
\mathcal{R}=\frac{\mathcal{B}(\Lambda_b\rightarrow \Lambda_c K)}{\mathcal{B}(\Lambda_b\rightarrow \Lambda_c \pi)},
\end{eqnarray}
for which
the uncertainty due to hadronic effects cancels to a large extent.
The ratio can be used to test factorization of amplitudes.
Since the two decays have similar topology and kinematic properties,
in the limit of $U$-spin symmetry, the ratio $\mathcal{R}$ is dominated by the ratio of the relevant CKM matrix
elements $|V_{us}/V_{ud}|^2\approx 0.054$.
After including the decay constants $f_{\pi}$ and  $f_{K}$, the ratio increases to about  0.081,
which is approach to our prediction 0.071 without the nonfactorization contributions.
When the  nonfactorization effects are considered,
the ratio changes slightly because the decay rates of the two modes increase to be a comparable proportion.
The LHCb collaboration has previously measured this ratio to be $0.0731\pm 0.0016\pm 0.0016$~\cite{prd89032001},
where the first uncertainty is statistical and the second is systematic,
which is compatible with our PQCD prediction.





\section{ conclusion}\label{sec:sum}
In this study,
we have performed a systematic
analysis of  the color-allowed baryonic decays $\Lambda_b\rightarrow \Lambda_c P$ with
 $P$ denotes either kaon or pion  in the framework of PQCD.
 With the previous well determined leading-twist LCDAs of baryons, we calculated the branching ratios and up-down asymmetry parameters
 including both the factorizable and nonfactorizable contributions.
 At the PQCD leading order in $\alpha_s$, a nonleptonic two-body baryon decay process requires at least two hard gluon exchanges.
 All the possible Feynman diagrams can be classified into five topological types, namely $T$, $C$, $E$, $B$, and $G$.
 14 diagrams of $T$-type and 2 diagrams of $G$-type contribute to the factorizable amplitude, while others enter into the nonfactorizable one.
 It is observed that $T$-type diagrams dominates and the contributions from other four type diagrams are small, totaling less than $10\%$.
 The three-gluon contributions are between those of $W$-emission  and  $W$-exchange diagrams.
 The concerned processes are dominated by the factorizable contribution,
 but the nonfactorizable contributions are also important and cannot be ignored.
 Their constructive interference enhances the decay branching ratios.

The predicted two baryonic transition form factors $f_1$ and $g_1$ at maximum recoil
are close to those in other theoretical methods within errors,
but larger than the previous PQCD calculation due to
the different treatment of the charm quark and charmed baryon masses.
It is found that the nonzero mass difference could numerically enhance the branching ratios of the two-body decay
and improve comparison with experiments.
The predicted branching ratios of $\Lambda_b\rightarrow \Lambda_c \pi$ and $\Lambda_b\rightarrow \Lambda_c K$
are  $6.7^{+4.1}_{-3.8}\times 10^{-3}$ and $5.0^{+4.0}_{-3.0}\times 10^{-4}$, respectively,
which are generally larger than other theoretical
results due to the involved important nonfactorizable effects.
The obtained ratio of kaon to pion decay rates is accessible to the experiments by LHCb.
As expected in the heavy quark limit and the soft meson limit,
the asymmetry parameter is predicted to be -1,
which indicate the $V-A$ nature of the weak current and maximum parity violation.
Since only the tree operator contributes to the decays,
the direct $CP$ asymmetries are absent naturally.

We discussed theoretical uncertainties arising from
the hadronic parameters in baryon LCDAs, the heavy charm quark and charmed baryon masses, and the hard scales.
The branching ratios suffer a large theoretical uncertainties, 
whereas the up-down asymmetries  are less sensitive to these parameters.
It should be emphasized that here only the leading-twist baryon LCDA are considered,
while other higher-twist contributions are neglected within the accuracy of the current work,
which may give significant uncertainties.
The higher-twist contributions should be included to improve the precision of theoretical predictions in the future.
As a case study with rough estimation,
our results may provide useful and reference information for the future application of the PQCD approach to the heavy baryon decays.

It would be interesting to extend our analysis to another two charmful nonleptonic decays
 $\Lambda_b\rightarrow\Lambda_c^+ D^-$ and $\Lambda_b\rightarrow\Lambda_c^+ D_s^-$,
in which it would involve also contributions from penguin diagrams.
The interference between tree and penguin amplitudes will produce the $CP$ asymmetry.
In particular, the Cabibbo-favored mode $\Lambda_b\rightarrow\Lambda_c^+ D_s^-$ was observed
with a large decay branching ratio of the order $10^{-2}$.
The $CP$ asymmetries in such processes are worthy to be explored in the future.

\begin{acknowledgments}
We would like to acknowledge Hsiang-nan Li, Yu-Ming Wang, Yue-Long Shen, and Ya Li for helpful discussions.
This work is supported by National Natural Science Foundation
of China under Grants No. 12075086 and  No. 11605060 and the Natural Science Foundation of Hebei Province
under Grants No.A2021209002 and  No.A2019209449.
\end{acknowledgments}

\begin{appendix}
\section{Factorization formulas for the $T$-type diagrams}\label{sec:wil}
\begin{table}[!htbh]
\caption{The Wilson coefficients $a$ and the virtualities of the internal gluon $t_{A,B}$
and quark $t_{C,D}$ for each diagram $T_{ij}$ in Fig.~\ref{fig:FeynmanT}. }
\label{tab:wil}
\begin{tabular}[t]{lccccc}
\hline\hline
$R_{ij}$ & $\frac{a}{8}$        & $\frac{t_A}{M^2}$ & $\frac{t_B}{M^2}$ & $\frac{t_C}{M^2}$          & $\frac{t_D}{M^2}$                 \\ \hline
$T_{a1}$ & $C_2+\frac{1}{3}C_1$ & $x_3x_3'$         & $(1-x_1)(1-x_1')$ & $(1-x_1)x_3'$              & $1-x_1'$                          \\
$T_{a2}$ & $C_2+\frac{1}{3}C_1$ & $x_3x_3'$         & $(1-x_1)(1-x_1')$ & $(1-x_1')x_3$              & $1-x_1'$                          \\
$T_{a3}$ & $C_2+\frac{1}{3}C_1$ & $x_3x_3'$         & $x_2x_2'$         & $x_2+x_3'-x_2x_3'$         & $1-x_1'$                          \\
$T_{a5}$ & $C_2+\frac{1}{3}C_1$ & $x_3x_3'$         & $x_2x_2'$         & $x_2'+x_3-x_2'x_3$         & $r_c^2+(1-x_2')(x_3-r^2)$         \\
$T_{a6}$ & $\frac{1}{3}C_1$     & $x_3x_3'$         & $x_2x_2'$         & $x_2'+x_3-x_2'x_3$         & $x_3'(x_3-(1-r^2)y)$              \\
$T_{a7}$ & $\frac{1}{3}C_1$     & $x_3x_3'$         & $x_2x_2'$         & $x_2'+x_3-x_2'x_3$         & $x_3'(y-1+x_3+r^2(1-y))$          \\
$T_{b1}$ & $C_2+\frac{1}{3}C_1$ & $x_3x_3'$         & $(1-x_1)(1-x_1')$ & $(1-x_1)x_3'$              & $1+r_c^2-r^2-x_1$                 \\
$T_{b2}$ & $C_2+\frac{1}{3}C_1$ & $x_3x_3'$         & $(1-x_1)(1-x_1')$ & $(1-x_1')x_3$              & $1+r_c^2-r^2-x_1$                 \\
$T_{b4}$ & $C_2+\frac{1}{3}C_1$ & $x_3x_3'$         & $x_2x_2'$         & $1+r_c^2-r^2-x_1$          & $r_c^2+x_2+r^2(x_3'-1)-x_2x_3'$   \\
$T_{b6}$ & $\frac{1}{3}C_1$     & $x_3x_3'$         & $x_2x_2'$         & $x_3'(x_3+(r^2-1)y)$       & $r_c^2+x_2+r^2(x_3'-1)-x_2x_3'$   \\
$T_{b7}$ & $\frac{1}{3}C_1$     & $x_3x_3'$         & $x_2x_2'$         & $x_3'(x_3-1-r^2(y-1)+y)$   & $r_c^2+x_2+r^2(x_3'-1)-x_2x_3'$   \\
$T_{c1}$ & $\frac{1}{3}C_1$     & $x_3x_3'$         & $(1-x_1)(1-x_1')$ & $(1-x_1)x_3'$              & $(1-x_1')(1-x_1-y+r^2y)$          \\
$T_{c2}$ & $\frac{1}{3}C_1$     & $x_3x_3'$         & $(1-x_1)(1-x_1')$ & $(1-x_1')x_3$              & $(1-x_1')(1-x_1-y+r^2y)$          \\
$T_{c5}$ & $\frac{1}{3}C_1-\frac{1}{4}C_2$ & $x_3x_3'$ & $x_2x_2'$      & $x_3'(x_3+(r^2-1)y)$       & $(1-x_1')(1-x_1-y+r^2y)$          \\
$T_{c7}$ & $\frac{1}{3}C_1-\frac{1}{4}C_2$ & $x_3x_3'$ & $x_2x_2'$      & $x_3'(x_3-1-r^2(y-1)+y)$   & $x_2'(x_2+(r^2-1)y)$              \\
$T_{d1}$ & $\frac{1}{3}C_1$     & $x_3x_3'$         & $(1-x_1)(1-x_1')$ & $(1-x_1)x_3'$              & $(x_1'-1)(x_1+r^2(y-1)-y)$        \\
$T_{d2}$ & $\frac{1}{3}C_1$     & $x_3x_3'$         & $(1-x_1)(1-x_1')$ & $(1-x_1')x_3$              & $(x_1'-1)(x_1+r^2(y-1)-y)$        \\
$T_{d6}$ & $\frac{1}{3}C_1-\frac{1}{4}C_2$ & $x_3x_3'$ & $x_2x_2'$      & $(x_1'-1)(x_1+r^2(y-1)-y)$ & $x_2'(x_2-1-r^2(y-1)+y)$          \\
\hline\hline
\end{tabular}
\end{table}
In this Appendix we provide some details about the calculations of the $T$-type diagrams,
while those for the other type ones can be derived in a similar way.
It is interesting to note that some diagrams in Fig.~\ref{fig:FeynmanT}
are connected by an interchange of two light quarks.
We give one example for $T_{a1}$ as an illustration.
By interchanging $u\leftrightarrow d$, it turn to $T_{f1}$,
and thereby the amplitude of the $T_{f1}$ can be obtained from that of $T_{a1}$ by exchanging the momenta indices 2 and 3,
and the spinor indices $\beta$ and $\gamma$ simultaneously.
As can be seen in Eq.~(\ref{eq:wave}),
the baryon LCDA is symmetric in exchanging $x_2$ and $x_3$,
while the structure $[\cdots]_{\beta\gamma}$ are antisymmetric
under the permutation of the spinor indices.
When including two baryons in the initial and final states, their amplitudes remain the same.
Now, we collect all the diagrams  that satisfy the above relation as below
\begin{eqnarray}
T_{a1}&=&T_{f1},\quad T_{a2}=T_{e1},\quad T_{a3}=T_{a4},\quad T_{a5}=T_{b3},\quad T_{a6}=T_{c3},\quad T_{a7}=T_{d3}, \nonumber\\
T_{b1}&=&T_{f2},\quad T_{b2}=T_{e2},\quad T_{b4}=T_{b5},\quad T_{b6}=T_{c4},\quad T_{b7}=T_{d4},\quad T_{c1}=T_{f3}, \nonumber\\
T_{c2}&=&T_{e3},\quad T_{c5}=T_{c6},\quad T_{c7}=T_{d5},\quad T_{d1}=T_{f4},\quad T_{d2}=T_{e4},\quad T_{d6}=T_{d7}.
\end{eqnarray}
In the following we only present the results for diagrams on the left of the above equation.

The Wilson coefficients $a$, the virtualities of the internal gluon $t_{A,B}$
and quark $t_{C,D}$ for each diagram $T_{ij}$ in Fig.~\ref{fig:FeynmanT} are collected in Table~\ref{tab:wil}.
The $b$-space measures $[\mathcal{D}b]$ and the expressions of $\Omega$ are presented in Table~\ref{tab:bb},
where the auxiliary functions $h_l$ are defined as
\begin{eqnarray}
h_1(\textbf{b},A,B)&=&\frac{1}{16\pi^3}\int_0^1 dz \frac{\sqrt{\zeta_1}}{\sqrt{\eta_1}}
\left\{K_1(\sqrt{\zeta_1\eta_1})\theta(\eta_1)+\frac{\pi}{2}[N_1(\sqrt{-\zeta_1\eta_1})-i J_1(\sqrt{-\zeta_1\eta_1})]\theta(-\eta_1)\right\}, \nonumber\\
\zeta_1(\textbf{b})&=&|\textbf{b}|^2, \nonumber\\
\eta_1(A,B)&=&Az+B\bar{z}, \\
h_2(\textbf{b}_A,\textbf{b}_B,A,B,C)&=&\frac{1}{32\pi^3}\int_0^1\frac{dz_1dz_2}{z_1\bar{z}_1} \frac{\sqrt{\zeta_2}}{\sqrt{\eta_2}}
\left\{K_1(\sqrt{\zeta_2\eta_2})\theta(\eta_2)+\frac{\pi}{2}[N_1(\sqrt{-\zeta_2\eta_2})-i J_1(\sqrt{-\zeta_2\eta_2})]\theta(-\eta_2)\right\}, \nonumber\\
\zeta_2(\textbf{b}_A,\textbf{b}_B)&=&|\textbf{b}_A-z_1 \textbf{b}_B|^2+\frac{z_1\bar{z}_1}{z_2}|\textbf{b}_B|^2, \nonumber\\
\eta_2(A,B,C)&=&A\bar{z}_2+\frac{z_2}{z_1\bar{z}_1}[B\bar{z}_1+Cz_1], \\
h_3(\textbf{b}_A,\textbf{b}_B,\textbf{b}_C,A,B,C,D)&=&\frac{1}{64\pi^3}\int_0^1\frac{dz_1dz_2dz_3}{z_1\bar{z}_1z_2\bar{z}_2} \frac{\sqrt{\zeta_3}}{\sqrt{\eta_3}}
\left\{K_1(\sqrt{\zeta_3\eta_3})\theta(\eta_3)+\frac{\pi}{2}[N_1(\sqrt{-\zeta_3\eta_3})-i J_1(\sqrt{-\zeta_3\eta_3})]\theta(-\eta_3)\right\}, \nonumber\\
\zeta_3(\textbf{b}_A,\textbf{b}_B,\textbf{b}_C)&=&|\textbf{b}_A-\textbf{b}_Bz_2-\textbf{b}_Cz_1\bar{z}_2|^2+
\frac{z_2\bar{z}_2}{z_3}|\textbf{b}_B-\textbf{b}_Cz_1|^2+\frac{z_1\bar{z}_1z_2\bar{z}_2}{z_2z_3}|\textbf{b}_C|^2, \nonumber\\
\eta_3(A,B,C,D)&=&A\bar{z}_3+\frac{z_3}{z_2\bar{z}_2}\left\{B\bar{z}_2+\frac{z_2}{z_1\bar{z}_1}[C\bar{z}_1+Dz_1]\right\},
\end{eqnarray}
with the Bessel functions $K_{0,1}$, $N_1$, and $J_1$ and the Feynman parameters $z_l$ and $\bar {z}_l=1-z_l$.
We do not repeat
some useful Fourier integration formulas for the derivation of the above functions, which could be found in Refs.~\cite{prd80034011,prd65074030}.
The hard amplitudes $H_{R_{ij}}$ are gathered in the  Table~\ref{tab:amp}.


\begin{table}[!htbh]
\caption{The expressions of $[\mathcal{D}b]$ and $\Omega_{R_{ij}}$ for the $T$-type diagrams.}
\label{tab:bb}
\begin{tabular}[t]{lccc}
\hline\hline
$R_{ij}$ & $[\mathcal{D}b]$ &$\Omega_{R_{ij}}$\\ \hline
$T_{a1}$ & $\int d^2 \textbf{b}_2d^2\textbf{b}_3d^2\textbf{b}'_2d^2\textbf{b}'_3$
 & $\frac{1}{(2\pi)^4}K_0(\sqrt{t_A}|\textbf{b}_2-\textbf{b}_3|)K_0(\sqrt{t_B}|\textbf{b}_3+\textbf{b}'_3-\textbf{b}'_2|)
 K_0(\sqrt{t_C}|\textbf{b}_2-\textbf{b}_3+\textbf{b}'_2-\textbf{b}'_3|)K_0(\sqrt{t_D}|\textbf{b}_3+\textbf{b}'_3|)$ \\

 $T_{a2}$ & $\int d^2 \textbf{b}_2d^2\textbf{b}_3d^2\textbf{b}'_2d^2\textbf{b}'_3$
 & $\frac{1}{(2\pi)^4}K_0(\sqrt{t_A}|\textbf{b}_2-\textbf{b}_3|)K_0(\sqrt{t_B}|\textbf{b}_2|)
 K_0(\sqrt{t_C}|\textbf{b}'_2-\textbf{b}'_3+\textbf{b}'_2-\textbf{b}'_3|)K_0(\sqrt{t_D}|\textbf{b}_3+\textbf{b}'_3|)$ \\

$T_{a3}$ & $\int d^2 \textbf{b}_3 d^2\textbf{b}'_2 d^2\textbf{b}'_3$
 &$K_0(\sqrt{t_A}|\textbf{b}_3|)h_2(-\textbf{b}'_2,-\textbf{b}_3-\textbf{b}'_3,t_B,t_C,t_D)$ \\

$T_{a5}$ & $\int d^2 \textbf{b}_2 d^2\textbf{b}'_2 d^2\textbf{b}'_3$
 &$K_0(\sqrt{t_A}|\textbf{b}'_3|)K_0(\sqrt{t_B}|\textbf{b}_2|)h_1(-\textbf{b}_2-\textbf{b}'_2,t_C,t_D)$ \\

$T_{a6}$ & $\int d^2 \textbf{b}_qd^2\textbf{b}_2d^2\textbf{b}'_2d^2\textbf{b}'_3$
 & $\frac{1}{(2\pi)^4}K_0(\sqrt{t_A}|\textbf{b}_q-\textbf{b}'_3|)K_0(\sqrt{t_B}|\textbf{b}_2|)
 K_0(\sqrt{t_C}|\textbf{b}_2+\textbf{b}'_2|)K_0(\sqrt{t_D}|\textbf{b}_q|)$ \\

$T_{a7}$ & $\int d^2 \textbf{b}_qd^2\textbf{b}_2d^2\textbf{b}'_2d^2\textbf{b}'_3$
 & $\frac{1}{(2\pi)^4}K_0(\sqrt{t_A}|\textbf{b}_q+\textbf{b}'_3|)K_0(\sqrt{t_B}|\textbf{b}_2|)
 K_0(\sqrt{t_C}|\textbf{b}_2+\textbf{b}'_2|)K_0(\sqrt{t_D}|\textbf{b}_q|)$ \\

$T_{b1}$ & $\int d^2 \textbf{b}_2d^2\textbf{b}_3d^2\textbf{b}'_2d^2\textbf{b}'_3$
 & $\frac{1}{(2\pi)^4}K_0(\sqrt{t_A}|\textbf{b}_2-\textbf{b}_3|)K_0(\sqrt{t_B}|\textbf{b}'_2|)
 K_0(\sqrt{t_C}|\textbf{b}_2-\textbf{b}_3+\textbf{b}'_2-\textbf{b}'_3|)K_0(\sqrt{t_D}|\textbf{b}_3+\textbf{b}'_3|)$ \\

$T_{b2}$ & $\int d^2 \textbf{b}_2d^2\textbf{b}_3d^2\textbf{b}'_2d^2\textbf{b}'_3$
 & $\frac{1}{(2\pi)^4}K_0(\sqrt{t_A}|\textbf{b}'_2-\textbf{b}'_3|)K_0(\sqrt{t_B}|\textbf{b}_2-\textbf{b}_3-\textbf{b}'_3|)
 K_0(\sqrt{t_C}|\textbf{b}_2-\textbf{b}_3+\textbf{b}'_2-\textbf{b}'_3|)K_0(\sqrt{t_D}|\textbf{b}_3+\textbf{b}'_3|)$ \\

$T_{b4}$ & $\int d^2 \textbf{b}_3 d^2\textbf{b}'_2 d^2\textbf{b}'_3$
 &$K_0(\sqrt{t_B}|\textbf{b}'_2|)h_2(-\textbf{b}'_3,-\textbf{b}_3-\textbf{b}'_3,t_A,t_C,t_D)$ \\

$T_{b6}$ & $\int d^2 \textbf{b}_qd^2\textbf{b}_3d^2\textbf{b}'_2d^2\textbf{b}'_3$
 & $\frac{1}{(2\pi)^4}K_0(\sqrt{t_A}|\textbf{b}_q+\textbf{b}_3|)K_0(\sqrt{t_B}|\textbf{b}'_2|)
 K_0(\sqrt{t_C}|\textbf{b}_q|)K_0(\sqrt{t_D}|\textbf{b}_3+\textbf{b}'_3|)$ \\

$T_{b7}$ & $\int d^2 \textbf{b}_qd^2\textbf{b}_3d^2\textbf{b}'_2d^2\textbf{b}'_3$
 & $\frac{1}{(2\pi)^4}K_0(\sqrt{t_A}|\textbf{b}_3-\textbf{b}_q|)K_0(\sqrt{t_B}|\textbf{b}'_2|)
 K_0(\sqrt{t_C}|\textbf{b}_q|)K_0(\sqrt{t_D}|\textbf{b}_3+\textbf{b}'_3|)$ \\

$T_{c1}$ & $\int d^2 \textbf{b}_qd^2\textbf{b}_2d^2\textbf{b}'_2d^2\textbf{b}'_3$
 & $\frac{1}{(2\pi)^4}K_0(\sqrt{t_A}|\textbf{b}_2+\textbf{b}'_3|)K_0(\sqrt{t_B}|\textbf{b}_q-\textbf{b}'_2|)
 K_0(\sqrt{t_C}|\textbf{b}_2+\textbf{b}'_2|)K_0(\sqrt{t_D}|\textbf{b}_q|)$ \\

$T_{c2}$ & $\int d^2 \textbf{b}_qd^2\textbf{b}_2d^2\textbf{b}'_2d^2\textbf{b}'_3$
 & $\frac{1}{(2\pi)^4}K_0(\sqrt{t_A}|\textbf{b}'_2-\textbf{b}'_3|)K_0(\sqrt{t_B}|\textbf{b}_q+\textbf{b}_2|)
 K_0(\sqrt{t_C}|\textbf{b}_2+\textbf{b}'_2|)K_0(\sqrt{t_D}|\textbf{b}_q|)$ \\

$T_{c5}$ & $\int d^2 \textbf{b}_q d^2\textbf{b}_2 d^2\textbf{b}_3$
 &$K_0(\sqrt{t_A}|\textbf{b}_3+\textbf{b}_q|)h_2(\textbf{b}_2+\textbf{b}_q,\textbf{b}_q,t_B,t_D,t_C)$ \\

$T_{c7}$ & $\int d^2 \textbf{b}_q d^2\textbf{b}'_2 d^2\textbf{b}'_3$
 &$h_3(\textbf{b}_q+\textbf{b}'_3,\textbf{b}'_2,\textbf{b}_q,t_A,t_B,t_C,t_D)$ \\

$T_{d1}$ & $\int d^2 \textbf{b}_qd^2\textbf{b}_2d^2\textbf{b}'_2d^2\textbf{b}'_3$
 & $\frac{1}{(2\pi)^4}K_0(\sqrt{t_A}|\textbf{b}_2+\textbf{b}'_3|)K_0(\sqrt{t_B}|\textbf{b}_q+\textbf{b}'_2|)
 K_0(\sqrt{t_C}|\textbf{b}_2+\textbf{b}'_2|)K_0(\sqrt{t_D}|\textbf{b}_q|)$ \\

$T_{d2}$ & $\int d^2 \textbf{b}_qd^2\textbf{b}_2d^2\textbf{b}'_2d^2\textbf{b}'_3$
 & $\frac{1}{(2\pi)^4}K_0(\sqrt{t_A}|\textbf{b}'_2-\textbf{b}'_3|)K_0(\sqrt{t_B}|\textbf{b}_2-\textbf{b}_q|)
 K_0(\sqrt{t_C}|\textbf{b}_2+\textbf{b}'_2|)K_0(\sqrt{t_D}|\textbf{b}_q|)$ \\

$T_{d6}$ & $\int d^2 \textbf{b}_q d^2\textbf{b}'_2 d^2\textbf{b}'_3$
 &$K_0(\sqrt{t_B}|\textbf{b}'_2+\textbf{b}_q|)h_2(-\textbf{b}_q,-\textbf{b}_q-\textbf{b}'_3,t_A,t_C,t_D)$ \\
\hline\hline
\end{tabular}
\end{table}

\begin{table}[H]
\center
\caption{The expressions of $H^{S,P}_{R_{ij}}$ for the $T$-type diagrams.}
\newcommand{\tabincell}[2]{\begin{tabular}{@{}#1@{}}#2\end{tabular}}
\label{tab:amp}
\begin{tabular}[t]{lcc}
\hline\hline
$R_{ij}$&$\frac{H^S}{16(1-r)\phi^A(y)M^5}$&$\frac{H^P}{16(1+r)\phi^A(y)M^5}$\\\hline
$T_{a1}$&$-2((r-2)r(x_1-1)x_1'+(1-2r)x_3')$&$-2((r-2)r(x_1-1)x_1'+(1-2r)x_3')$\\
$T_{a2}$&$2(r-2)r-2((1-2r)x_3+(r-2)r)x_1'$&$2(r-2)r-2((1-2r)x_3+(r-2)r)x_1'$\\
$T_{a3}$&\tabincell{c}{$x_3'(2r^2x_1'+(r-1)^2)$\\$+x_2((r-1)^2x_1'+2)-2(r+1)(rx_1'+1)$}&\tabincell{c}{$x_3'(2r^2x_1'+(r-1)^2)$\\$+x_2((r-1)^2x_1'+2)-2(r+1)(rx_1'+1)$}\\
$T_{a5}$&\tabincell{c}{$x_2'(-(r^3+r)x_2'+r(r(r+r_c+2)$\\$+2r_c+3)-r_c)-2r(r+1)(r_c+1)$}&\tabincell{c}{$x_2'(r(r(-r+3r_c+6)-2r_c+1)+r((r$\\$-4)r+1)x_2'+r_c)-2r(r+1)(r_c+1)$}\\
$T_{a6}$&$-2((r^2-1)y+x_3)(rx_2'-r-1)-r(r-1)^2x_2'x_3'$&$(r-1)^2rx_2'x_3'-2((r^2-1)y+x_3)(rx_2'-r-1)$\\
$T_{a7}$&$(r-1)^2x_3x_3'-2((r^2-1)(y-1)-x_3)(rx_2'-r-1)$&$(r-1)^2x_3x_3'-2((r^2-1)(y-1)-x_3)(rx_2'-r-1)$\\
$T_{b1}$&$-2r((r_c-2rr_c)x_3'+(r-2)x_1-r+2)$&$-2r((r_c-2rr_c)x_3'+(r-2)x_1-r+2)$\\
$T_{b2}$&$2(r-2)r^2r_c(x_2'+x_3')+(4r-2)x_3$&$2(r-2)r^2r_c(x_2'+x_3')+(4r-2)x_3$\\
$T_{b4}$&\tabincell{c}{$-2(r+1)r^2r_c-((r-2)r+3)r^2$\\$+((r-4)r+1)r_c^2+r((r-1)^2r_c$\\$+2r)x_3'+x_2(2rr_c+(r-1)^2)$}&\tabincell{c}{$-2(r+1)r^2r_c-((r-2)r+3)r^2$\\$+((r-4)r+1)r_c^2+r((r-1)^2r_c$\\$+2r)x_3'+x_2(2rr_c+(r-1)^2)$}\\
$T_{b6}$&\tabincell{c}{$(r-1)(x_3'((r-1)r(r-r_c)$\\$-(r-1)x_2-2r(r+1)y)+2r(r$\\$+1)(r_c+1)y)+2rx_3(r_c-x_3'+1)$}&\tabincell{c}{$(r-1)(x_3'((r-1)r(r-r_c)$\\$-(r-1)x_2-2r(r+1)y)+2r(r$\\$+1)(r_c+1)y)+2rx_3(r_c-x_3'+1)$}\\
$T_{b7}$&\tabincell{c}{$-(2rx_3(r_c-x_3'+1)-(r-1)(x_3'(r($\\$-2(r+1)y+r+r_c+3)+(r-1)rx_3'$\\$-r_c)+2r(r+1)(r_c+1)(y-1)))$}&\tabincell{c}{$(r-1)(x_3'(-r(2(r+1)y-3r+r_c-1)$\\$-(r-1)rx_3'+r_c)+2r(r+1)(r_c$\\$+1)(y-1))-2rx_3(r_c-x_3'+1)$}\\
$T_{c1}$&$-2(2r-1)x_3'((r^2-1)y-x_1+1)$&$-2(2r-1)x_3'((r^2-1)y-x_1+1)$\\
$T_{c2}$&$2(r-2)r(x_1'-1)((r^2-1)y-x_1+1)$&$2(r-2)r(x_1'-1)((r^2-1)y-x_1+1)$\\
$T_{c5}$&\tabincell{c}{$-((r^2-1)y+x_2+x_3)(2(r^2$\\$-1)y+(r-1)^2(-x_3')+2x_3)$}&\tabincell{c}{$-((r^2-1)y+x_2+x_3)(2(r^2$\\$-1)y+(r-1)^2(-x_3')+2x_3)$}\\
$T_{c7}$&\tabincell{c}{$(r-1)^2rx_2'x_3'-2((r^2-1)y$\\$+x_2)((r^2-1)(y-1)-x_3)$}&\tabincell{c}{$-2((r^2-1)y+x_2)((r^2-1)(y$\\$-1)-x_3)-r(r-1)^2x_2'x_3'$}\\
$T_{d1}$&$-2(2r-1)x_3'((r^2-1)(y-1)-x_2-x_3)$&$-2(2r-1)x_3'((r^2-1)(y-1)-x_2-x_3)$\\
$T_{d2}$&$2(r-2)r(x_1'-1)(r^2(y-1)+x_1-y)$&$2(r-2)r(x_1'-1)(r^2(y-1)+x_1-y)$\\
$T_{d6}$&\tabincell{c}{$-((r^2-1)(y-1)-x_2-x_3)(2(r^2$\\$-1)(y-1)+(r-1)^2x_2'-2x_2)$}&\tabincell{c}{$-((r^2-1)(y-1)-x_2-x_3)(2(r^2$\\$-1)(y-1)+(r-1)^2x_2'-2x_2)$}\\
\hline\hline
\end{tabular}
\end{table}

\end{appendix}

\end{document}